\documentclass[conference]{IEEEtran}
\IEEEoverridecommandlockouts
\usepackage{amsmath,amssymb,amsfonts}
\usepackage{algorithmic}
\usepackage{graphicx}
\usepackage{textcomp}
\usepackage{xcolor}
\usepackage{caption}
\usepackage{url}
\usepackage{svg}
\usepackage{tabularx,booktabs}
\usepackage[colorlinks=true, linkcolor=blue, citecolor=blue, urlcolor=blue]{hyperref}
\usepackage{cite}
\def\BibTeX{{\rm B\kern-.05em{\sc i\kern-.025em b}\kern-.08em
    T\kern-.1667em\lower.7ex\hbox{E}\kern-.125emX}}
\begin{document}

\title{VIVID: A Novel Approach to Remediation Prioritization in Static Application Security Testing (SAST)
}

\author{\IEEEauthorblockN{Naeem Budhwani}
\IEEEauthorblockA{
\textit{Accenture}\\
Montreal, Canada \\
naeem.budhwani@gmail.com}
\and
\IEEEauthorblockN{Mohammad Faghani}
\IEEEauthorblockA{
\textit{Accenture}\\
Toronto, Canada \\
mrfaghani@gmail.com}
\and
\IEEEauthorblockN{Hayden Richard}
\IEEEauthorblockA{
\textit{Accenture}\\
Nashville, USA \\
haydenrichard411@gmail.com}
}

\maketitle

\begin{abstract}
Static Application Security Testing (SAST) enables organizations to detect vulnerabilities in code early; however, major SAST platforms do not include visual aids and present little insight on correlations between tainted data chains. We propose VIVID - Vulnerability Information Via Data flow - a novel method to extract and consume SAST insights, which is to graph the application's vulnerability data flows (VDFs) and carry out graph theory analysis on the resulting VDF directed graph. Nine metrics were assessed to evaluate their effectiveness in analyzing the VDF graphs of deliberately insecure web applications. These metrics include 3 centrality metrics, 2 structural metrics, PageRank, in-degree, out-degree, and cross-clique connectivity. We present simulations that find that out-degree, betweenness centrality, in-eigenvector centrality, and cross-clique connectivity were found to be associated with files exhibiting high vulnerability traffic, making them refactoring candidates where input sanitization may have been missed. Meanwhile, out-eigenvector centrality, PageRank, and in-degree were found to be associated with nodes enabling vulnerability flow and sinks, but not necessarily where input validation should be placed. This is a novel method to automatically provide development teams an evidence-based prioritized list of files to embed security controls into, informed by vulnerability propagation patterns in the application architecture.
\end{abstract}


\section{Introduction}
Eight of the top ten data breaches of 2023 were related to application attack surfaces \cite{crowdstrike2024}. This demands robust application-layer countermeasures, including active Static Application Security Testing (SAST) scanning. SAST performs analyses \cite{bychkov2022} including:

\begin{enumerate}
    \item \textit{Data flow analysis}: Follows the path of the data flow
    \item \textit{Control flow analysis}: Compares the application control flow on execution with known-secure code flow patterns
    \item \textit{Structural analysis}: Examines language-specific code structures for inconsistencies with best practices
    \item \textit{Semantic analysis}: Performs a simple search looking for known-insecure strings in the code base
    \item \textit{Configuration analysis}: Checks application configuration files against security best practices
\end{enumerate}

Our research focuses on (1) and (2), as these techniques lend themselves readily to graphical visualization. We begin by noticing that SAST tools identify data flow paths (e.g., function calls) where tainted data passes through without security controls (e.g., input validation and sanitization). At present, flow data can be gathered from the UI of some major SAST platforms and is used by development teams to remediate vulnerabilities \cite{synopsys_intro}. The potential for aggregating this data across vulnerabilities and running analytics on it remains untapped and is overlooked due to the cumbersome length of individual VDF data, which is typically excluded from standard reports \cite{sastissues}.

Our research involves gathering and graphing tainted data paths, which will be referred to as vulnerability data flows (VDFs). We define a VDF as the propagation of an untrusted value from a source collection point in code to a destination or intermediate destination in code. When VDFs are graphed, nodes in the resulting graph signify files through which tainted data passes, while graph edges refer to flowing tainted data. This tainted data is directional and the resulting graph is a directed graph (digraph), which means analysis can be carried out to identify features like feedback loops, temporal sequence, and flow imbalance.

The graph serves two productive uses:

\begin{enumerate}
    \item \textit{Rapid consumption}: Rather than looking at VDFs for each vulnerability in isolation from each other through a SAST user interface, the graph reveals nodes shared by multiple vulnerabilities. As a result, the files contributing most to the vulnerability of the application will be identified. Sections of the applications that are most vulnerable will also be able to be identified. 
    \item \textit{Generating insights using graph theory}: The graph of VDFs lends itself readily to measurement taking. We evaluate a variety of measurements in this paper, including centrality and other graph theory metrics, to determine their significance in an application security context.
\end{enumerate}

The contribution of this research is the demonstration that the application of graph theory on application security results, specifically a constructed VDF graph, is meaningful for development teams looking to prioritize remediation. 

This paper focuses on analyzing the VDF graph to optimize vulnerability reduction while minimizing the number of code commits. We test this experimentally by using graph theory metrics to identify files (i.e. graph nodes) involved in a maximum number of tainted data flows (i.e. paths in the graph) to retrieve a prioritized list of remediation and refactoring candidates. The remainder of this paper discusses existing work in the area and our contribution, followed by a discussion of our simulation set-up, results, and follow-on work. We demonstrate the insight that graph theory metrics can provide development teams so they can leverage vulnerability propagation patterns in application architectures when prioritizing tasks.

\section{Related Works}
While a study of the utility of graph theory metrics on VDF graphs has not been conducted to the best of our knowledge, there are a variety of approaches to analyze vulnerability data flows. Static taint analysis is one such approach and which tracks taints at the variable-level in code \cite{taintintro}. Much static taint analysis literature proposes domain-specific tools and methodologies such as Tripp et al.’s TAJ for Java \cite{TAJ} and Arzt et al.’s Flowdroid for Android \cite{flowdroid}, with the latter constructing a Taint Value Graph (TVG). Some modern approaches to taint analysis employ deep learning, with work including Niu et al’s approach for IoT \cite{IoTTaint} and Chow et al.’s Fluffy \cite{Fluffy} on top of GitHub’s CodeQL analysis framework \cite{codeql}.

Data flow analysis is another avenue to analyze vulnerability data flows. Treating programmatic data flows using formal theory was expounded by F.E. Allen and J. Cocke from the IBM Thomas J. Watson Research Center \cite{allen1976} in 1976. This paper laid out formal definitions for nodes and edges of data flow graphs. Researchers from the University of Colorado Boulder then built on IBM's work in the same year to perform anomaly calculations on the data flow graph \cite{fosdick1976}. The body of literature continues to treat data flow graphs as directed and accessible by graph theory. With the advent of SAST, Checkmarx has published work on hierarchical data-flow graphs \cite{pereira2023} in 2023. 

Moreover, some work has commented on the time-consuming nature of SAST output consumption and proposed visualizations via an interactive dashboard \cite{schreiber2021towards}. This visualization data draws from commits rather than data flows. Meanwhile, there has been a body of research establishing the usefulness of stack trace data to estimate the attack surface \cite{theisen2015} and to locate files responsible for specific vulnerabilities \cite{iyer2004}.

The area of software vulnerability assessment and prioritization has also been the subject of many surveys and reviews, including by Khan et al. (rule-based methods) \cite{khan2018}, Kritikos et al. (static analysis) \cite{kritikos2019}, Dissanayake et al. (socio-technical aspects) \cite{dissanayake2021}, and Le et al. (meta-survey) \cite{le2022}. These surveys show studies of data-driven techniques (including multi-layer perceptron, random forest, and linear SVM) to analyze data sources including ExploitDB \cite{bhatt2020}, NVD \cite{bullough2017}, dark web forums and markets \cite{almukaynizi2017}, and other open-source repositories \cite{xiao2018,jiang2020}. 

This paper contributes to the body of knowledge by offering graph theory analysis methods for vulnerability data flow graphs, with the aim of prioritizing vulnerable files to remediate and to generate intuitive visualizations. 

\section{Solution Design and Implementation Details}
VDFs are obtained with a SAST tool. VDFs are then taken from a Veracode API. The user can run VIVID locally on their machine after putting in the Veracode API key. Veracode expects VDF data in X format.
VIVID can be used with existing SAST tools. VIVID is a collection of scripts that can be executed over a command line interface. 

\section{Experiment Setup}
When choosing metrics to analyze the VDF graph, it was appealing to include both local metrics such as degree and global metrics such as centrality metrics. This would allow insights into both direct and indirect relationships and vulnerability flow within the application architecture. 

Nine metrics were chosen for analysis. These include 3 centrality metrics, 2 graph structural metrics, in-degree, out-degree, PageRank, and cross-clique connectivity. Refer to Table 1 for a comprehensive list of these metrics and their relevance in an application security context.

WebGoat (version 2023.8) and VeraDemo (version 2.1.1) were chosen as the application targets, both of which are deliberately insecure Java applications. The former is maintained by OWASP while the latter is maintained by Veracode, making them common targets for application security testing. Vulnerability data flows were then pulled from the API of a major SAST platform, from which the VDF graph was constructed. 

After constructing the graph using Gephi \cite{gephi} and GraphVis \cite{rossi2015}, an R script was developed and used to analyze the graphs against the 9 chosen metrics. Results and analysis discussion for each metric are presented below. 

\begin{figure}[htbp]
\centerline{\includegraphics{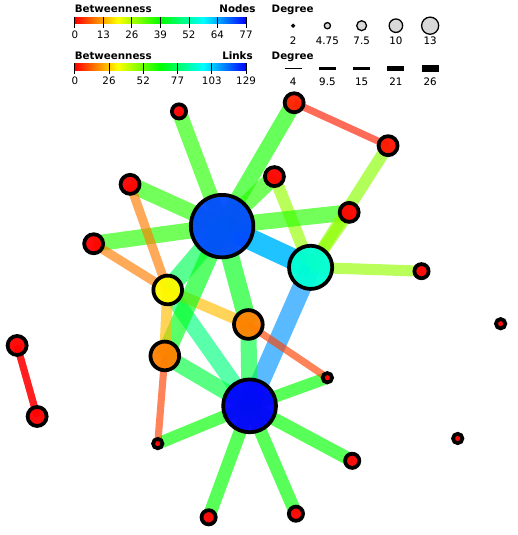}}
\caption{VDF visualization of VeraDemo}
\label{fig:VeraDemo}
\end{figure}

The VeraDemo VDF graph is presented in Figure \ref{fig:VeraDemo}.

The VDF graph of WebGoat is shown below. The graph includes many vulnerability islands in the architecture, implying that most vulnerabilities are not data-flow related or that the data flow is restricted to a single file. It is also seen that the large blue node in the middle corresponds to WebGoatUser.java, which contains constructors and serves up information such as the user role, username, or password on request.

\begin{figure}[htbp] 
	\includegraphics[width=\linewidth]{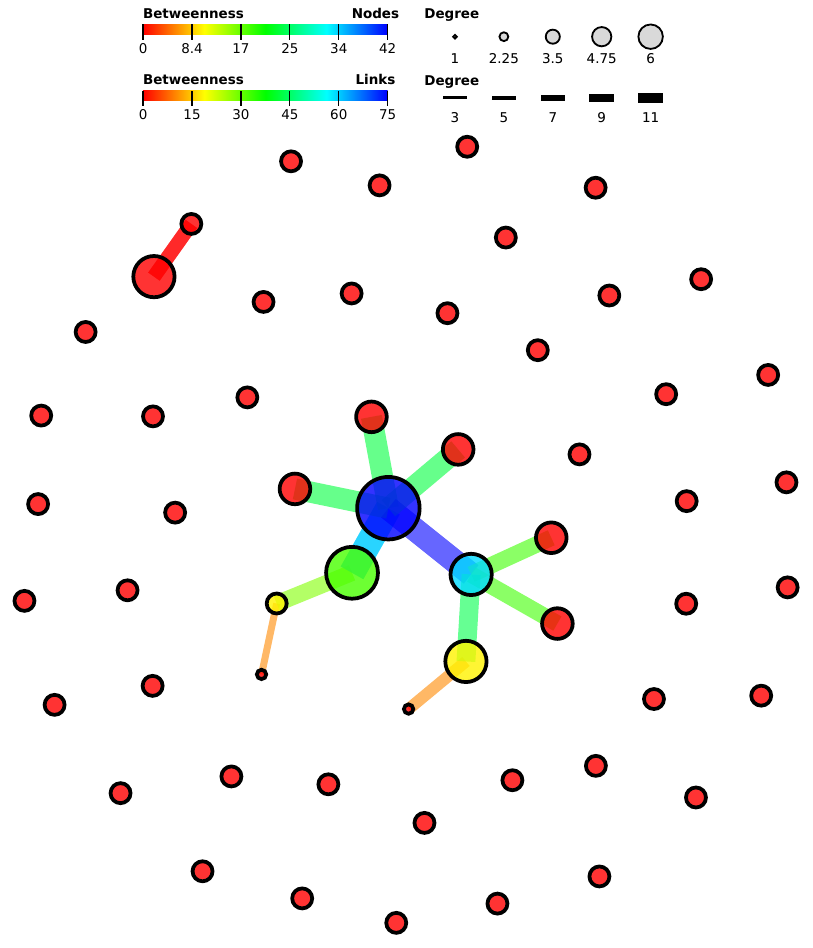}
	\caption{VDF graph for WebGoat v2023.8}
        \label{fig:WebGoat VDF Graph}
\end{figure}

A manual analysis was conducted on WebGoat code, finding that WebGoatUser employs a Model-View-Controller (MVC) architecture where any data validation is typically done as soon as the data is received, which is in the controller file. In other words, the file flagged as vulnerable by the SAST tool may not necessarily be the file where security controls such as validation or sanitization should be introduced.

To measure the success of the metrics in analyzing the VDF graph, we recall our objective to pinpoint files through which tainted data frequently traverses, enabling development teams to prioritize the integration of input validation, sanitization, and other security metrics. In so doing, we enumerated a list of 5 files of interest in WebGoat and VeraDemo from a manual code review and measure the success of the metrics in terms of the measure's capture of these files of interest.

All scripts are included in the referenced GitHub repository for reproduction.

\begin{table*}[h!] 
\centering
\begin{tabularx}{\textwidth}{|>{\hsize=.3\hsize}X|>{\hsize=1.3\hsize}X|>{\hsize=1.4\hsize}X|}
\hline
\textbf{Measure} & \textbf{Definition} & \textbf{Significance in an Application Security Context}\\ 
\hline
Out- and In- Eigenvector Centrality & \[ \mathbf{v}_i = \frac{1}{\lambda} \sum_{j=1}^{N} A_{ij} \mathbf{v}_j \] denotes the out-eigenvector centrality of file \( i \), where \( A_{ij} \) indicates a tainted data path from file \( i \) to file \( j \), and \( \lambda \) quantifies the maximum potential "traffic" or exerted influence in tainted data path network \( A_{ji} \) is used for in-eigenvector centrality. & Identifies files that propagate vulnerability chains and contribute to the impact magnitude of vulnerability spread. Pathways of nodes with high eigenvector centrality provide a clear picture of the predominant pathways of influence across the network. Development teams may choose to prioritize the remediation of vulnerabilities for which data passes through files with high eigenvector centralities whose compromise would entail a significant blast radius.  \\
\hline
Substructure Entropy & \[ H(v) = -\sum_{u \in V} p(u|v) \log p(u|v) \] where \( p(u|v) \) is the probability of tainted data flowing from file \( v \) to file \( u \). & Unusual occurrence, suggesting the relative level of refactoring effort required to remediate vulnerabilities. \\
\hline
Modularity & \[ Q = \frac{1}{2|E|} \sum_{vw} \left[ A_{vw} - \frac{\text{deg}(v) \text{deg}(w)}{2|E|} \right] \delta(c_v, c_w) \] where \( |E| \) is the number of tainted data paths and \( \delta \) being the Kronecker delta function, checking membership. & Number of communities, indicating clustered vulnerabilities or interrelated security issues. \\
\hline
Betweenness Centrality & \[ C_B(v) = \sum_{s\neq v \neq t} \frac{\sigma_{st}(v)}{\sigma_{st}} \] where \( \sigma_{st} \) is total number of shortest tainted paths from file \( s \) to file \( t \) and \( \sigma_{st}(v) \) denotes paths through file \( v \). & Identifies bridges and bottlenecks in the application flow where vulnerabilities can have a higher spread or likelihood. A file with high betweenness centrality indicates that it is frequently encountered during the most efficient (shortest) routes that tainted data might take as it propagates through the system. \\
\hline
PageRank & \[ PR(u) = \frac{1-d}{N} + d \sum_{v \in B_u} \frac{PR(v)}{L(v)} \] with \( d \) as damping (\textasciitilde0.85), \( N \) total files, \( B_u \) files sending tainted data to file \( u \), and \( L(v) \) files receiving tainted data from file \( v \). & Identifies common sinks in vulnerability data flows. Optionally, PageRank weights may be assigned from the vulnerability's severity. \\
    \hline
     In-Degree &  \begin{equation*} D^-(v) = |{(u,v) \mid (u,v) \in E}| \end{equation*} where \( u \) and \( v \) are files and \( E \) is the edge set.  & Identifies functionally critical application areas where input validation is imperative. These include I/O interfaces, centralized databases, API endpoints, shared utilities, middleware, caches, inter-process buffers, and temp storage. High in-degree indicates essential roles, numerous interactions, and potential failure points. \\
    \hline
     Out-Degree &  \begin{equation*}
D^+(v) = |{(v,u) \mid (v,u) \in E}|
\end{equation*} where \( u \) and \( v \) are files and \( E \) is the edge set. & Identifies vulnerablity disseminators. These include I/O outputs, database queries, API responses, library usages, middleware dispatches, cache updates, inter-process signals, and temp transfers. High out-degree suggests broad impacts, multiple dependencies, and influence spread.  \\
    \hline
     Cross-Clique Connectivity & \( X(v) \) is the number of cliques to which node \( v \) belongs. &  Identifies highly cross-connected nodes, showing a tight coupling of vulnerabilities and node involvement in numerous vulnerability clusters. \\
    \hline

\end{tabularx}
\caption{Description of graph theory metrics in an application security setting}
\end{table*}

\section{Results and Discussion}
Simulations on 2 deliberately insecure web applications were run and whose results are shown in the below radar graphs, where metrics are shown around the circumference of the graph. The graph shows how each of the 5 files of interest were ranked by the respective metric, where higher-ranked files are closer to the centre of the graph.

\begin{figure}[htbp]
\centerline{\includegraphics[width=\linewidth]{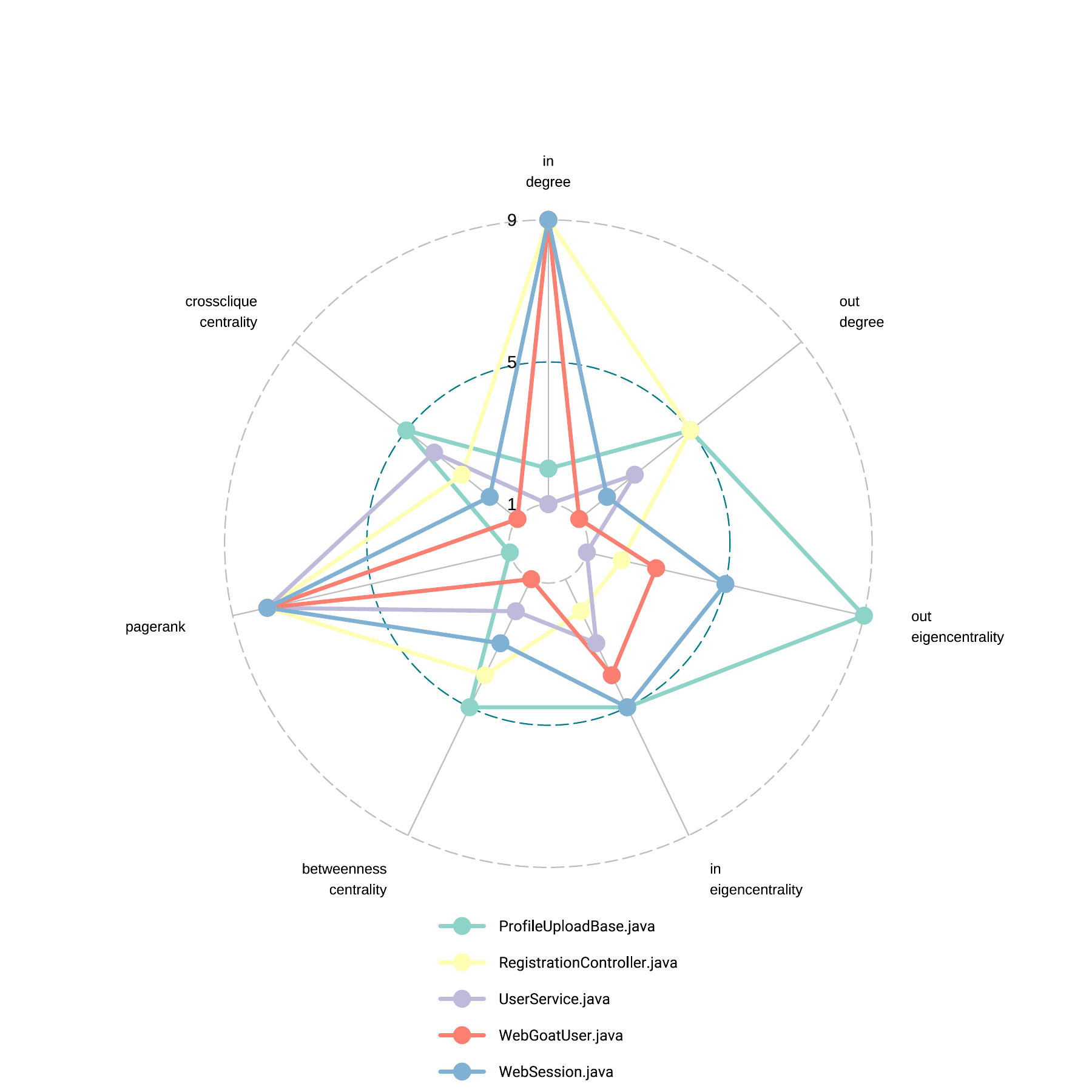}}
\caption{Simulation results for WebGoat}
\label{fig:WebGoat Results}
\end{figure}

Figure \ref{fig:WebGoat Results} shows the WebGoat results, where in-degree and PageRank fail to capture several files of interest and out-eigencentrality captured some files of interest in their top-ranked files. Cross-clique connectivity, betweenness centrality, out-eigencentrality, and in-eigencentrality captured all files of interest in their top 5 rankings.

\begin{figure}[htbp]
\centerline{\includegraphics[width=\linewidth]{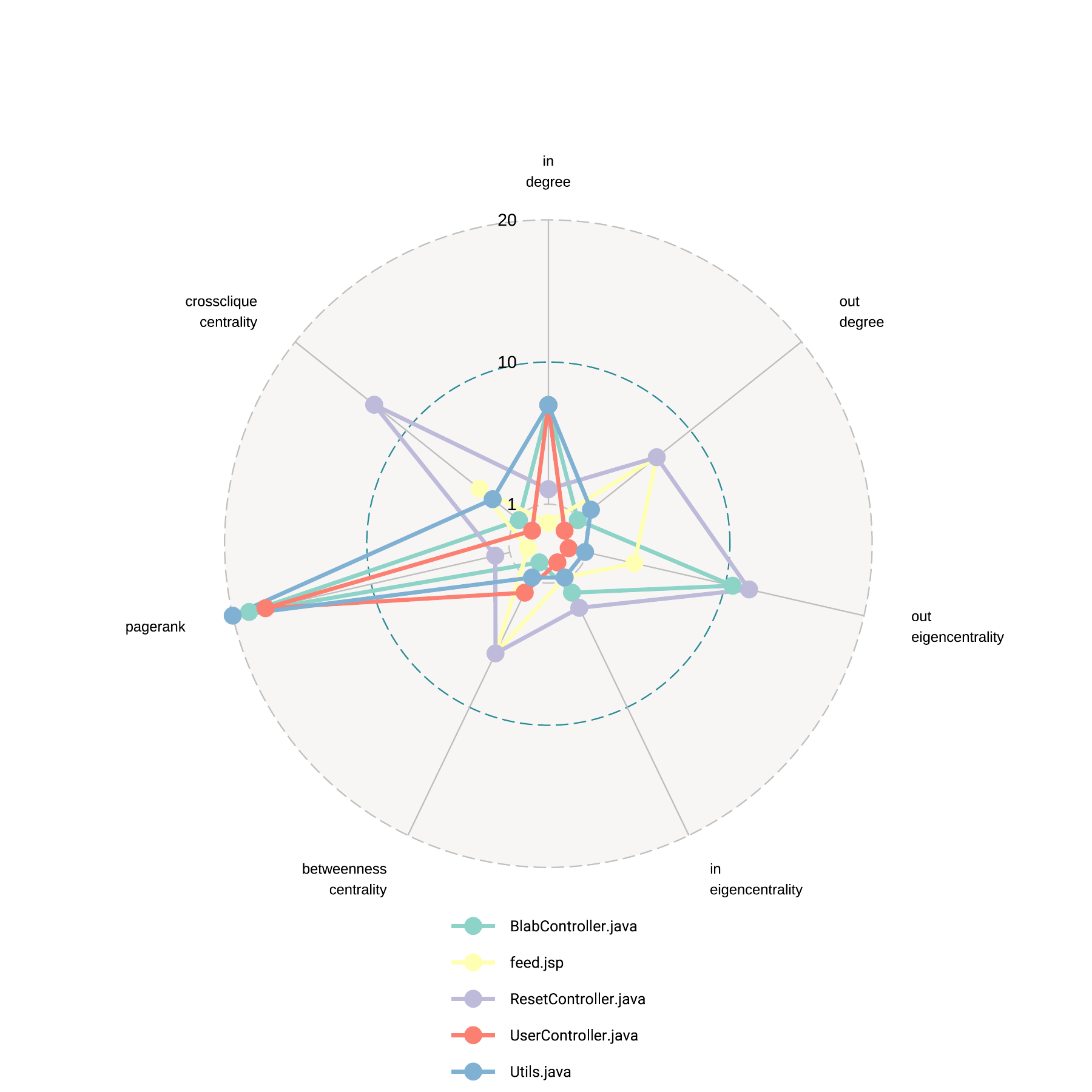}}
\caption{Simulation results for VeraDemo}
\label{fig:VeraDemo Results}
\end{figure}

Results from the VeraDemo simulation are shown in Figure \ref{fig:VeraDemo Results}. The results here echo WebGoat results in that PageRank fails to capture key of interest and that out-eigencentrality captures only some files of interest in its top 10. Note that the axis is re-scaled for VeraDemo to account for a larger number of nodes in the VDF graph.

\subsection{Betweenness Centrality}
Betweenness centrality is commonly used in network analysis. While existing tools like Sonargraph validate the entire application architecture by constructing module dependency graphs and cyclicity graphs \cite{sonargraph2023}, there are no tools to model application vulnerabilities in the context of application architecture.

Armed with the concept of bridges in its mathematical definition, using betweenness centrality directly aligns with the stated objective of identifying files where tainted data frequently passes through. By focusing remediation on these nodes which correspond to bridges or chokepoints, development teams can address a disproportionately larger number of vulnerabilities, minimizing code commits while maximizing vulnerability reduction. 

In the simulation of WebGoat, betweenness centrality determined the WebGoatUser.java to be the node with the highest betweenness centrality of 21. This node is seen in the centre of the graph, shown in Figure \ref{fig:WebGoat Results}. Its surrounding nodes at betweenness centrality 16, 12, and 9. All remaining nodes have a betweenness centrality of 0.

The analysis shows that the WebGoatUser.java, which is the node with the highest degree (6) in the graph, is called and so accessible to its neighboring nodes. Placing a generic input validation function here would be a quick win as the function can be called by all its neighboring nodes and so the fix would mitigate many vulnerabilities. The simulation on VeraDemo confirms that betweenness centrality identifies high-value targets for input validation. 

Simulation results for VeraDemo seen in \ref{fig:VeraDemo Results} show that the BlabController.java has the highest betweenness centrality. Identifying a controller as a remediation target in an MVC application is a good sign. Interestingly, BlabController.java had the highest betweenness centrality whereas all other centrality metrics ranked UserController.java as the highest in their respective measure. This suggests that betweenness centrality provides key information on vulnerability remediation targets differentiated from other centrality metrics and which should be taken into account in a weighted formula that outputs a prioritized list of files to remediate.

\subsection{Eigenvector Centrality}
In the realm of networks, some nodes are not influential merely because they have many connections; they are influential because they connect to other influential nodes. The idea behind eigenvector centrality is to quantify this recursive notion of influence.

Nodes or sections of the VDF graph with high eigenvector centrality scores would be flagged as high-risk due to their potential cascading impact on the larger system. As a result, this may lead to an automatic generation of refactoring candidates.

Results on the WebGoat VDF show that 12 nodes have a non-zero out-eigenvector centrality. UserService.java has the highest out-eigenvector centrality (that is, of value 1), while the RegistrationController.java has an out-eigenvector centrality of 0.79. In-eigenvector centrality captures files including UserForm.java and ProfileUploadRetrieval.java in its top 5 rankings which are useful places to add in input validation. 

In-eigencentrality out-performs out-eigencentrality for the VeraDemo simulation as the former picks up on 4 controller files in its top 5 rankings. Recalling that controllers are files of interest to insert security controls, this metric should be weighted higher when developing a formula to prioritize refactoring and insertion of security controls.

\subsection{Modurality and Substructure Entropy}

A modularity analysis was also carried out to visually represent clusters within the VDF graph. The cluster\textunderscore walktrap was used as a modularity measure since it is appropriate for directed graphs. 

\begin{figure}[htbp] 
	\includegraphics[width=\linewidth]{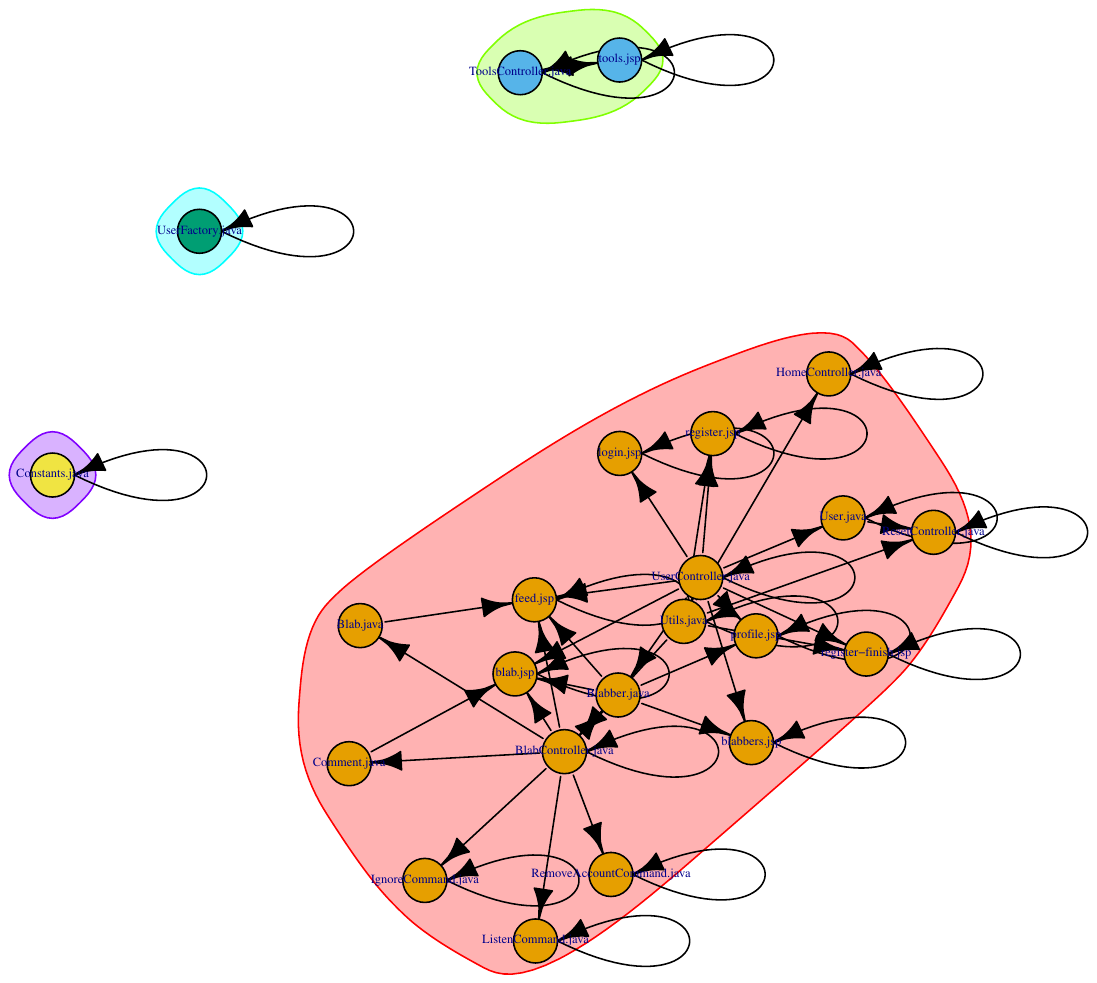}
	\caption{Modurality plot for WebGoat v2023.8}
        \label{fig:WebGoat Modularity Plot}
\end{figure}

Figure \ref{fig:WebGoat Results} shows the modularity plot of WebGoat. The large section of entropy 2.2 can be seen in the red background while the section of entropy 1 and 2 nodes can be seen on its upper left. The remainder of the nodes were not of sufficient degree to be included in the plot.

To be useful for development teams in prioritizing segments of the application, entropy calculations can be made on each of these vulnerability islands found in VDF graph.

Substructure entropy can be used for this purpose and offers immediate use cases for identifying refactoring candidates and third-party security contexts. It is a measure that captures the unpredictability or randomness of substructures within the VDF graph.

A higher entropy value suggests a diverse vulnerability propagation pattern, whereas a lower value might indicate recurring vulnerability patterns. This entropy measure provides a lens through which we can better understand the intricacy and patterns of vulnerabilities in applications. We enumerate two use cases for this measure:

\emph{1. Identifying refactoring candidates: }Acquirers can leverage substructure entropy on the VDF graph to quantify and compare the refactoring effort required in different regions of the application code. Segments of the target software asset’s code showing higher entropy in their VDF graph merit special attention, as they emerge as prime candidates for refactoring. 

Sections of target software asset code with high substructure entropy represent potential risks and liabilities associated with the target’s software asset. This is because these high-entropy code sections hinder the software's security maintainability and security resilience. 

\emph{2. Third-party dependencies \& External dependencies: } 96\% of codebases surveyed contain open-source software, according to Synopsys' 2023 Report on Open Source Security and Risk Analysis. More jarring is that 76\% of code by volume was open-source \cite{synopsysopensource2023}. Software assets not only integrate but largely lie on top of third-party libraries and dependencies. Given the complexity of how first-party (1P) and third-party (3P) code intermingle and are tightly woven, upgrading outdated 3P software may require refactoring efforts. So, estimating the size of refactoring efforts involving 3P is of interest to development teams and the business. 

At the time of writing, the authors were not able to find a tool that quantifies refactoring efforts involved in upgrading 3P libraries and packages to a secure version. Our suggested approach for this is to create a vulnerability data flow (VDF) graph on application code and rank software sections based on their entropy. Sections with higher entropy scores would need more intensive refactoring efforts.

When calculating substructure entropy using the R script provided in the GitHub repository, analysis minimizes noise by excluding nodes of degree two which are self-connections (loop) and not connected to any other node. In the case of WebGoat, there are 2 connected segments of entropy 2.2 and 1, which house 12 nodes and 2 nodes respectively. Similarly in VeraDemo, the substructure entropy analysis identifies 2 vulnerability islands, housing 2 and 19 nodes. The entropy analysis reveals an entropy of 2.47 for the 19-node segment. 

Our analysis finds that these entropy values should be normalized so the segment's entropy can be interpreted as low, moderate, high, or very high. As such, we divide the found entropy by the maximum entropy given the degrees in the segment. The upper bounds for low, moderate, and high entropy were set as 0.25, 0.5, and 0.75 respectively. 

The entropy values support the case that a higher remediation effort is needed for the 2.2 entropy section of 12 nodes. Moreover, the separation of the 2 substructures in WebGoat are a result of them being located in different folder paths. 

\subsection{Cross-Clique Connectivity}
Cross-clique connectivity measures the propagation of information or disease in a graph \cite{faghani2013}, making it well-suited to VDF graph analysis. The cross-clique connectivity R package was used. This measure performed very well in the WebGoat simulation, as it captured 5 out of 5 (100\%) files of interest. Cross-clique connectivity's performance for VeraDemo was similar to the other centrality metrics, indicating a consensus on the key files of interest. 

\subsection{In-Degree and Out-Degree}
In and out-degree inform the extent of the in-flow and out-flow of tainted data flow respectively. Calculating files with the highest out-degree is equivalent to finding the list of files outputting the highest volumes of tainted data. This is important for developers to know where to place security controls including input validation. 

In regards to the performance of the out-degree measure shown in Figures \ref{fig:WebGoat Results} and \ref{fig:VeraDemo Results}, out-degree alone was able to capture 3 of 5 (60\%) major files of interest and was out-performed by centrality metrics such as cross-clique centrality which have a built-in understanding of influence in the larger graph. 

In-degree ranked only 2 of 5 (40\%) files of interest in its top 5 for WebGoat; however, it is a useful indicator of common sinks across multiple tainted data flow chains. Developers can insert post-operation checks or data integrity checks in files showing high in-degree. Nodes of high in-degree values should be provided as informational findings for development teams to insert these checks.

In-degree and out-degree performed identically for VeraDemo where the former ranked 2 of 5 (40\%) files of interest in its top 5 while the latter captured (60\%) in its top 5. 

\section{Conclusion}
Our paper demonstrates the fruitfulness of applying graph theory metrics to the VDF graph of an application. By harnessing vulnerability data flows (VDFs) and graphing them, we offer already data-saturated security departments an intuitive method to visualize application architecture.

We have shown that metrics including centrality metrics like cross-clique centrality can be applied out-of-the-box to VDF graphs, offering insight into prioritizing vulnerability remediations. Further, prioritization lists can then be generated automatically with further informational findings generated by in-degree file rankings. 

Moreover, we clarified how centrality metrics and other metrics shed light on vulnerability contexts. Namely, the metrics capture the interplay between vulnerabilities, as well as propagation patterns of tainted data, flagging specific files for attention. From this, we infer that remediation across a small subset of files can significantly curtail vulnerability spread. Based on our results, we advocate for development teams to hone in on files marked by high scores in these metrics when planning remediation priorities.

\subsection{Limitations \& Future Work}
Minimizing the number of code commits needed to maximize vulnerability reduction is but one of many avenues of analysis opened up by the use of graph theory on VDF graphs. A promising area of future work we are exploring is a points-based system, taking into account vulnerability severity. For example, higher points (i.e. priority) would be given to files where remediating one vulnerability (e.g., directory traversal) would mitigate multiple vulnerabilities, like command injection and XSS. 

There also exists potential in evaluating additional metrics and identifying synergies between metrics with the goal of developing an index or combination of weighted centrality metrics. To this end, node colouring or overlay can be used to infuse criticality data into the graph and make the weighted formula of prioritized files to remediate sensitive to vulnerability criticality.

In the way of experiential learning, we have included in our GitHub repository a script enabling transformed VDF data to be viewed in a Virtual Reality (VR) simulation, allowing analysts to explore and interact with static vulnerability data in immersive environments.

Looking forward, the vulnerability graph may be more granular, where each node is a function within a file rather than the file itself. This would allow the developer to be suggested a specific function where to put a input validator and sanitizer. A consideration on this approach is that many SAST vulnerabilities do not have a network effect; that is, they are contained within the same file, such as an insecure cipher suite vulnerability.

A noticeable gap exists in prioritizing remediation based on vulnerability severity. We look to examine the effectiveness of graph theory on graphs where weights correspond vulnerability severity and seek to publish our results in a follow-up study.

\bibliographystyle{IEEEtran}
\bibliography{references}

\end{document}